%% file: main.tex
\documentclass[runningheads]{llncs}
\usepackage{todonotes}
\usepackage{graphicx}
\usepackage{url}

\begin{document}
\title{On Dimensions of Plausibility for Narrative Information Access to Digital Libraries}
\titlerunning{On Dimensions of Plausibility for Narrative Information Access}
%
\author{Hermann Kroll \inst{1}\orcidID{0000-0001-9887-9276} \and
Niklas Mainzer\inst{1} \and Wolf-Tilo Balke\inst{1}\orcidID{0000-0002-5443-1215}}
\authorrunning{Kroll et al.}
%
\institute{Institute for Information Systems, TU Braunschweig, Braunschweig, Germany\\
\email{\{kroll,balke\}@ifis.cs.tu-bs.de and n.mainzer@tu-bs.de}}
\maketitle              
\begin{abstract}
Designing keyword-based access paths is a common practice in digital libraries. They are easy to use and accepted by users and come with moderate costs for content providers. However, users usually have to break down the search into pieces if they search for stories of interest that are more complex than searching for a few keywords. After searching for every piece one by one, information must then be reassembled manually. In previous work we recommended narrative information access, i.e., users can precisely state their information needs as graph patterns called narratives. Then a system takes a narrative and searches for evidence for each of its parts. If the whole query, i.e., every part, can be bound against data, the narrative is considered plausible and, thus, the query is answered. But is it as easy as that? In this work we perform case studies to analyze the process of making a given narrative plausible. Therefore, we summarize conceptual problems and challenges to face. Moreover, we contribute a set of dimensions that must be considered when realizing narrative information access in digital libraries.

\end{abstract}

\section{Introduction}
\input{01-Introduction}

\section{Case Studying Narrative Plausibility}
\input{02-CaseStudies}

\section{Concluding Discussion on Narrative Plausibility}
\input{03-NarrativePlausibility}

\bibliographystyle{splncs04}
\bibliography{references}

\end{document}

%% file: 01-Introduction.tex
Digital libraries are large-scale repositories containing a plethora of heterogeneous but well-curated data: texts, images, videos, research data, and many more.
Today, keyword-based search (on texts or respective metadata) forms the primary access path for collections: 
It is easy to use and thus a commonly accepted interface, and implementing a respective search index comes with acceptable costs for content providers. 
In addition, keyword queries can be evaluated against different data representations; See keyword-based searches in literature and books, in knowledge bases~\cite{elbassuni2011keywordsearchrdf,shuo2017keywordsearchkg}, and data set repositories~\cite{chapman2020datasetsearchsurvey}.
Although the advantage of keyword-based search is apparent, moving towards alternative access paths is heavily investigated, e.g., the biomedical database  SemMedDB~\cite{kilicoglu2012semmeddb} on top of Medline, the Open Knowledge Research Graph~\cite{jaradeh2019okrg}, or the Europeana as Linked Open Data~\cite{haslhofer2011europeana} and DBpedia~\cite{auer2007dbpedia}.
A good example is the current trend towards structured knowledge graphs: 
Here users can formulate their information needs as structured queries (e.g., via SPARQL interfaces) asking for a combination of knowledge particles (i.e., automatically joining pieces of information).

However, integrating different and heterogeneous sources into a single knowledge representation comes with new problems, such as the validity of fused information~\cite{kroll2020tpdl}, identifying and canonicalizing pieces of information (entity linkage), and practical extraction problems when mining knowledge from different sources; See~\cite{weikum2021machineknowledge} for a good overview. 
A remedy might be the use of \textit{narrative information access} to bypass the necessary integration of different sources.
In~\cite{kroll2020er}, \textbf{narratives} are defined by directed edge-labeled graph patterns involving entities (things and concepts), events (state or state changes), literals (values), and their respective relationships (properties, temporal, and causal predicates, etc.). 

Since the individual information particles within a narrative form proper subgraphs, narratives can be understood as forming a logical overlay on top of different knowledge repositories. 
A narrative query is then successfully answered if evidence for every single relationship of the narrative can be found in some collection's data.
This evidence forms so-called \textbf{narrative bindings} connecting each relationship against a concrete piece of information stated in some knowledge repository, i.e., any form of data storage (e.g., relational databases, knowledge graphs, data sets, etc.).
In previous works we formulated a conceptual model~\cite{kroll2020er} for narrative information access, discussed the binding process from a technical perspective~\cite{kroll2021text2story}, showed how queries can be processed over collections of scientific publications~\cite{kroll2021icadl}, and discussed conceptual problems like context compatibility when answering queries over independent knowledge bases~\cite{kroll2022jcdl}.

In contrast to fact-checking, where the task refers to deciding whether some statement is indeed a fact (and thus true) or is rejected because it has no evidence~\cite{uscinski2013factchecking,zhijiang2022factchecking}, narrative structures may be composed out of several statements. 
Moreover, we do not require a universally valid decision, i.e., based on a given ground truth (a theory in physics for example), we can make a narrative plausible, whereas we cannot make the narrative plausible on a different ground truth. 
While question answering and its methods~\cite{qioa2022bioQA,khot2017openieqa,zhau2020complexFQA,zhau2020complexFQA} are related to narrative information access, we see significant differences in the evaluation: Narrative information access considers contexts~\cite{kroll2022jcdl,kroll2020tpdl}, i.e., we do not fuse information from different contexts.
For a good example, consider claims from the biomedical domain: 
Combining claims made when treating mice with claims when treating humans can be a serious threat to the overall validity because no one guarantees that these claims should also be valid for humans then.

In summary, the key advantage of narrative information access is to make user-formulated graph patterns \textit{plausible} in valid contexts.
But beyond apparent technical problems and the problem of context compatibility in narrative information access, the notion of \textit{plausibility} has yet to be discussed in detail.
Our current model considers a narrative as plausible if evidence for every of its claims can be found.
But is it as easy as that?
In this paper we summarize the conceptual problems of our narrative query model.
Therefore, we perform case studies in different domains to find the basic building blocks for an actionable notion of plausibility.
Moreover, we open up a design space and discuss our notion of plausibility in detail by arguing on different evaluation semantics, the trustworthiness of sources, and the quality of bindings.

%% file: 02-CaseStudies.tex
\begin{figure}[t]
    \centering
    \includegraphics[trim=0.0cm 5cm 0.0cm 0.0cm, width=0.8\textwidth]{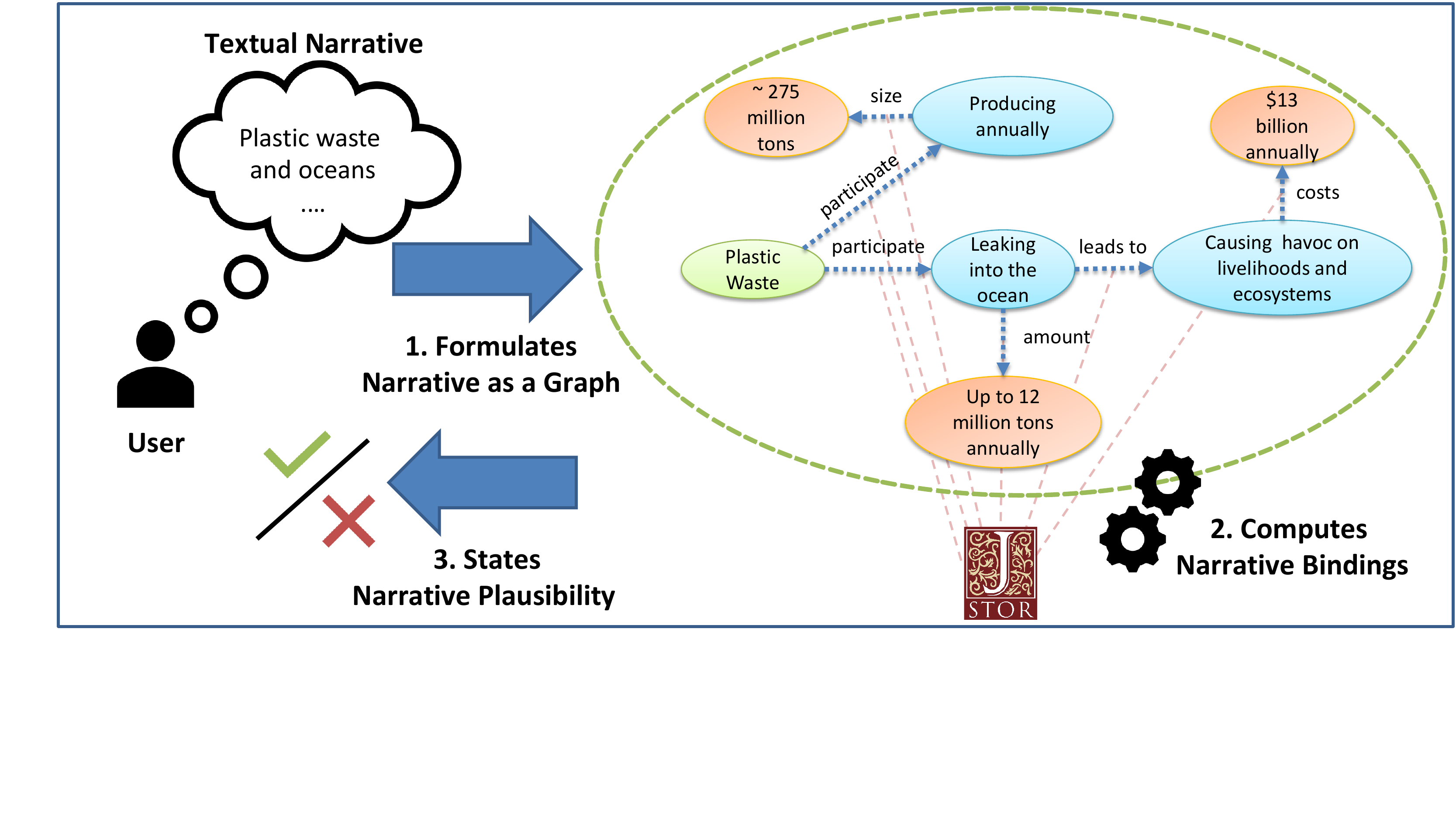}
    \caption{Making narratives plausible by computing bindings against knowledge repositories. Here our example Narrative \ref{plastic} is bound against the digital library JSTOR.}
    \label{fig:plasticextraction}
\end{figure}
In brief, a user models a narrative of interest as a graph structure. 
Then, a process searches for evidence for each relationship in a set of given knowledge repositories.
If we can find evidence for the whole narrative, i.e., every narrative's relationship, we consider it plausible~\cite{kroll2020er}.
The process is shown in Fig.~\ref{fig:plasticextraction}. 
Note that we may build different plausibility validation processes, which can vary in their included and excluded knowledge repositories for the binding computation, e.g., peer-reviewed scientific papers, fact-checked sources, or arbitrary websites.
Due to the lack of existing implementations for such a process, we performed manual case studies, i.e., we searched for evidence through keyword queries in knowledge repositories and evaluated the results manually.

\newtheorem{narrative}{Narrative}
\begin{narrative}
\label{plastic}
  Of the approximately 275 million metric tons of plastic waste produced annually, up to 12 million tons leak into oceans, wreaking havoc on livelihoods and ecosystems (CIEL, 2020). The result is an estimated \$13 billion in annual environmental damage to marine ecosystems.\emph{~\cite{narrative1} (See Fig.~\ref{fig:plasticextraction})}.
\end{narrative}

The central entity of our example is \textit{plastic waste}, around which the rest of the narrative revolves.
The narrative states two events directly relating to plastic waste, its \textit{annual production}, and that \textit{plastic waste is leaking into the ocean}. 
\textit{The havoc on livelihoods and ecosystems} is a direct consequence of \textit{ocean plastic} and represents the third event of our example.
Each event that we described is further connected to a literal: There are approximately \textit{275 million tons} of annual plastic waste production, up to\textit{ twelve million tons} leak into the ocean, and the havoc on livelihoods and ecosystems costs around \textit{13 billion dollars per year}.
Note that we may extract different graph representations of the same textual narrative but we discuss this transformation ambiguity in the next section. 

We next present a plausibility validation process based on the digital library JSTOR (\url{https://www.jstor.org/}).
One of the problems we faced in the beginning was that some relationships require more context from the narrative itself, e.g.,
\textit{Producing annually an amount of 275 million metric tons} does not include that it is about \textit{plastic waste}.
Either the model must include this detail in the event \textit{Producing annually}, or we must verify that the context of the production must match plastic waste.
Therefore, we started by evaluating that \textit{around 275 million metric tons of plastic waste are annually produced.}
We formulated different keyword queries to search for the given relationship.
We retrieved a source that claimed the 275 million metric ton estimation~\cite{jambeck}.
We found two more sources, one of which reported an annual plastic production of 381 million metric tons in 2015, of which 79 percent are considered waste~\cite{williams}.
We performed the calculation ourselves, and the result was around 300 million tonnes of plastic waste per annum~\cite{williams} or respectively 321 million metric tonnes per year~\cite{Rhodes2018PlasticPA}.

We then evaluated that \textit{up to 12 million tons of plastic waste are annually leaking into the ocean}.
\cite{Rhodes2018PlasticPA} reported that in 2010, approximately nine million tons of plastic were added to the ocean. 
\cite{williams} estimated that around 4.8 to 12.7 million tons of plastic entered the ocean in 2010. 
The plausibility validation seemed to confirm the narrative claim in this way.
We moved to the next relationship: \textit{The leaking of plastic into the ocean leads to havoc on livelihoods and ecosystems.} 
Even in 1987, \cite{azzarello} already reported an endangerment of marine environments caused by plastic in the ocean. 
A more recent article described dangers to the marine animal and plant life due to microplastic as well as the economic dangers~\cite{clark}.
For the last relation we searched evidence that the impact of plastic waste in oceans costs 13 billion dollars annually. 
\cite{Rhodes2018PlasticPA} reported 75 billion Dollar annual environmental costs originating from plastics. 
However, the source did not state how much of the total costs are connected to plastic waste in the ocean.
Apart from~\cite{Rhodes2018PlasticPA}, we could not find any other sources stating the actual costs.

\subsection{Summary of Other Case Studies}
We performed the plausibility validation for three additional narratives: 
1. Deforestation of the Amazonian Rainforest and forests in general, 
2. Claims about the connection between government measures during the Covid-19 pandemic and the number of Covid-related deaths, 
and 3. Global warming and the effects of carbon dioxide emissions on the climate.
For 3. we only considered fact-checked articles, whereas we allowed every website on the first result page of our Google Search for the 1. and 2. narrative. 

An interesting finding of the deforestation narrative occurred when we evaluated the percentage of the Earth's landmass covered by forests. 
The narrative stated a value of 30 percent. 
One of our retrieved sources confirmed this narrative claim by stating a value of 30.7 percent~\cite{worldbank}.
Another source, however, stated that forests cover 38 percent of the habitable land area~\cite{owidforestsanddeforestation}.
These values greatly differ and could be classified as counter-evidence because 30 percent could not be verified precisely. 
However, we noticed that~\cite{worldbank} base their value on the total land area, while~\cite{owidforestsanddeforestation} makes this calculation based on the habitable land area, explaining the difference in these values.

We performed a selection of our approaches multiple times over a period of several weeks to observe how the results would change over time. 
We retrieved the global warming narrative from~\cite{nasa}. 
When we initially accessed this source, the narrative reported an average surface temperature of 1.18 degrees Celsius on Earth.
However, over time this value has been changed to one degree Celsius, which is less specific than the initial estimate. 
While performing the initial plausibility validation, we noticed various factors determining Earth's average surface temperature changes, such as the data collection methods and the point in time used as a reference. 
Various sources described warming of around one degree, but we could not precisely confirm the 1.18 degrees Celsius.
The fact that NASA changed that value confirms our assumption that there is too much uncertainty about the data to state such a specific temperature change.

%% file: 03-NarrativePlausibility.tex
In the following we summarize the main findings from our previous case studies.

\textit{Confirmation Bias.}
Confirmation bias describes ”the seeking or interpreting of evidence in ways that are partial to existing beliefs, expectations, or a hypothesis at hand”~\cite{nickerson}.
As this definition suggests, bias can already occur in the search for information, and its effects highly depend on how the plausibility validation process searches for evidence. 
For instance, evaluating \textit{plastic waste causes 13 billion dollars annually in environmental costs by leaking into the ocean} could lead to other results than searching without a concrete cost value.

\textit{Narrative's Context.}
The narrative itself spans a context that must be considered when binding its relationships individually.
For instance, searching for \textit{an amount of leakage} without \textit{plastic waste} may result invalid results.
Similarly, evaluating the \textit{average surface temperature} without a concrete year may also yield results that do not match the user's indented context.

\textit{Finding the Narrative's Source.}
Imagine a narrative extracted from a study reports scientifically disputed findings.
We would assume such a narrative to be evaluated as implausible in a plausibility validation process. 
However, the process will likely find the study's sources again -- even if they are disputed.

\textit{Ambiguity of Narrative Extraction.}
Narratives can be interpreted differently since they make various claims, usually without explaining exactly how everything is meant.
Without an objective understanding of what the narrative is exactly stating, this will affect the narrative model extracted from the actual textual narrative. 
Consequently, the plausibility validation process computes narrative bindings for a selected instance of the possible interpretations, making the conclusion about narrative plausibility dependent on the chosen narrative model and, therefore, on the chosen interpretation of the narrative. 
A plausibility validation process based on another interpretation of the same narrative might result in a different conclusion about the narrative's plausibility.

\textit{Evidence.}
We observed three issues when searching for evidence: 
1. Indirect Evidence: The process may not find direct but rather indirect evidence (e.g., a certain percentage of the annual plastic production), which requires further computations. 
2. Counter-Evidence: We may find counter-evidence, i.e., evidence that contradicts a narrative's relationship. 
3. Absence of Evidence: Not finding evidence is always connected to uncertainty since we cannot be sure if evidence does not exist or if we simply did not find it.
All issues are especially relevant if we utilize error-prone NLP and retrieval methods~\cite{Gandrabur,blatz-etal-2004-confidence}.

\textit{Trustworthiness of Sources.}
When computing bindings against different types of sources, the credibility of sources must be considered.
On the one hand, we may utilize Provenance information that has already been curated~\cite{ProvO,prov-overview,carroll2005namedgraphs}.
Provenance allows knowledge curators to precisely describe how a certain piece of information was created or obtained. 
If we bind against that certain piece, we can then utilize its Provenance.
On the other hand, we may integrate tests like the CRAAP test~\cite{blakeslee} or automated heuristics, e.g., scientific publications before websites, highly influential over low cited publications, etc.
Note that we are aware of highly influential papers that have been retracted but handling such cases in digital libraries is a research field on its own~\cite{jodi2020knowledgemaintenance}.
But if we do not have curated information available, we must think about alternatives here.

\subsection{Dimensions of Narrative Plausibility}
The contribution of this work is the introduction of \textbf{dimensions} influencing narrative plausibility. When implementing a validation process in a digital library, the designers must decide how the following criteria should be handled.

\paragraph{\textbf{Narrative Structure.}}
We have seen that transforming a story of interest (a user's query) into a graph representation can be challenging and may yield several different representations. 
Moreover, each representation might end up in a different result (plausible/not plausible) which could confuse users.
Do we consider the exact narrative, or do we allow a few derivations (e.g., relaxing the query a bit)?
What can users then expect if a narrative is considered plausible?

\paragraph{\textbf{Validation Approach.}}
Which knowledge repositories are considered for the binding computation? 
Which methods are used to compute the actual bindings?
What do these methods guarantee? 
And thus, what could a user then expect?

\paragraph{\textbf{Types of Evidence.}}
Beyond suitable evidence we may face the \textit{indirect}, \textit{missing}, and \textit{counter-evidence} in practice.
What should we do if we cannot find evidence for all of a narrative's relationships? 
Is the narrative simply considered as \textit{not plausible}? 
Or should we instead show the user how much of the narrative is plausible and which parts lack evidence? 
Do we consider counter-evidence? 
And if we do, what does the existence of counter-evidence then mean, e.g., that the narrative is not plausible?
As we have argued before, there is a fine line between evidence not exactly confirming a relationship and counter-evidence.

\paragraph{\textbf{Confidence of Bindings.}}
How trustworthy do the sources have to be? 
How should this trustworthiness be determined or measured? 
How is the quality of the actual bindings from a technical perspective, e.g., the confidence of a retrieval process or similarity measures for values?
Should we use thresholds to consider trustworthy sources and high-quality bindings only?
Which evaluation strategy do we select, e.g., the trustworthiness of sources over binding quality?
Do we average both \textit{values} into an overall confidence?